\documentclass[nofootinbib]{revtex4}

\pdfoutput=1
\usepackage{graphicx}

\usepackage{newtxtext}       %
\usepackage{newtxmath}       
\usepackage{amsmath}
\usepackage{framed}

\begin{document}

\title{Phase and amplitude description of complex oscillatory patterns in reaction-diffusion systems}
\author{Hiroya Nakao}
\affiliation{Department of Systems and Control Engineering, Tokyo Institute of Technology, Tokyo 152-8552, Japan}

\begin{abstract}
Spontaneous rhythmic oscillations are widely observed in various real-world systems.
In particular, biological rhythms, which typically arise via synchronization of many self-oscillatory cells, often play important functional roles in living systems.
One of the standard theoretical methods for analyzing synchronization dynamics of oscillatory systems is the phase reduction for weakly perturbed limit-cycle oscillators, which allows us to simplify nonlinear dynamical models exhibiting stable limit-cycle oscillations to a simple one-dimensional phase equation.
Recently, the classical phase reduction method has been generalized to infinite-dimensional oscillatory systems such as spatially extended systems and time-delayed systems, and also to include amplitude degrees of freedom representing deviations of the system state from the unperturbed limit cycle.
In this chapter, we discuss the method of phase-amplitude reduction for spatially extended reaction-diffusion systems exhibiting stable oscillatory patterns.
As an application, we analyze entrainment of a reaction-diffusion system exhibiting limit-cycle oscillations by an optimized periodic forcing and additional feedback stabilization.
\end{abstract}

\maketitle

\section{Introduction}
\label{sec:1}

There are abundant examples of spontaneous rhythmic oscillations in living systems, ranging from microscopic oscillations of cardiac cells and spiking neurons to macroscopic oscillations of heartbeats and brainwaves~\cite{Strogatz,Winfree,Glass,Tass,Pikovsky,Ermentrout1}.
In many cases, macroscopic oscillations result from synchronized collective dynamics of many microscopic cells and play essentially important functional roles in the survival of living systems.
Regular rhythmic dynamics of the cells are typically modeled as limit-cycle oscillations in nonlinear dynamical systems.
A representative example of such dynamical systems is the Hodgkin-Huxley model of spiking neurons, which is a 4-dimensional ordinary differential equations (ODEs) with complex nonlinear terms representing the dynamics of the membrane potential and channel variables~\cite{Ermentrout1}.

Due to nonlinearity, 
analytical solutions of limit cycles are rarely available,
requiring approximate theoretical approaches to their synchronization dynamics.
The phase reduction~\cite{Ermentrout1,Kuramoto,Hoppensteadt,Kuramoto2,Ermentrout2,Nakao0,Stankovski,Monga,Pietras,Brown} is a classical and standard theoretical method for analyzing weakly perturbed limit-cycle oscillators, which approximately describes the oscillator state by using only a single phase value and simplifies the multidimensional nonlinear dynamical equations of the oscillator to a one-dimensional phase equation. It has been successfully used in analyzing, e.g., nonlinear waves and collective oscillations in populations of weakly coupled limit-cycle oscillators.

Recently, the method of phase reduction 
has been extended in several ways. In particular, (i) generalization 
to infinite-dimensional systems such as partial differential equations (PDEs)~\cite{Nakao} and delay-differential equations~\cite{Kotani,Novicenko} and (ii) inclusion of 
amplitude degrees of freedom representing
deviations 
of the system state
from the unperturbed limit cycle~\cite{Wedgwood,Wilson,Shirasaka} have been formulated.
The first extension is important in analyzing collective oscillations that arise e.g. in spatially extended populations of 
dynamical units described by PDEs.
The second extension, which gives reduced amplitude equations in addition to the phase equation, is relatively recent even for ODEs,
but it is necessary for describing transient relaxation dynamics of the perturbed system state to the limit cycle and can be used e.g. for stabilizing the oscillations by introducing feedback control of the amplitudes.

In this chapter, we formulate the method of phase and amplitude reduction for stable oscillatory patterns arising in spatially extended reaction-diffusion systems.

\section{Phase-amplitude reduction of 
limit-cycle oscillators}\label{sec:2}

We first review the method of phase-amplitude reduction~\cite{Wedgwood,Wilson,Shirasaka} for 
finite-dimensional limit-cycle oscillators described by ODEs of the form
\begin{align}
\dot{\bf X}(t) = {\bf F}({\bf X}(t)),
\label{eq1}
\end{align}
where ${\bf X}(t) \in {\mathbb R}^n$ is a $n$-dimensional oscillator state at time $t$ and ${\bf F} : {\mathbb R}^n \to {\mathbb R}^n$ is a sufficiently smooth vector field representing the system dynamics.
We assume that Eq.~(\ref{eq1}) has an exponentially stable limit-cycle solution ${\bf X}_0(t)$ of natural period $T$ and frequency $\omega = 2\pi / T$, satisfying ${\bf X}_0(t+T) = {\bf X}_0(t)$.
We denote this limit-cycle attractor as $\chi$ and its basin of attraction as $B \subseteq {\mathbb R}^n$.

\begin{figure}[t]
\includegraphics[width=0.4\hsize]{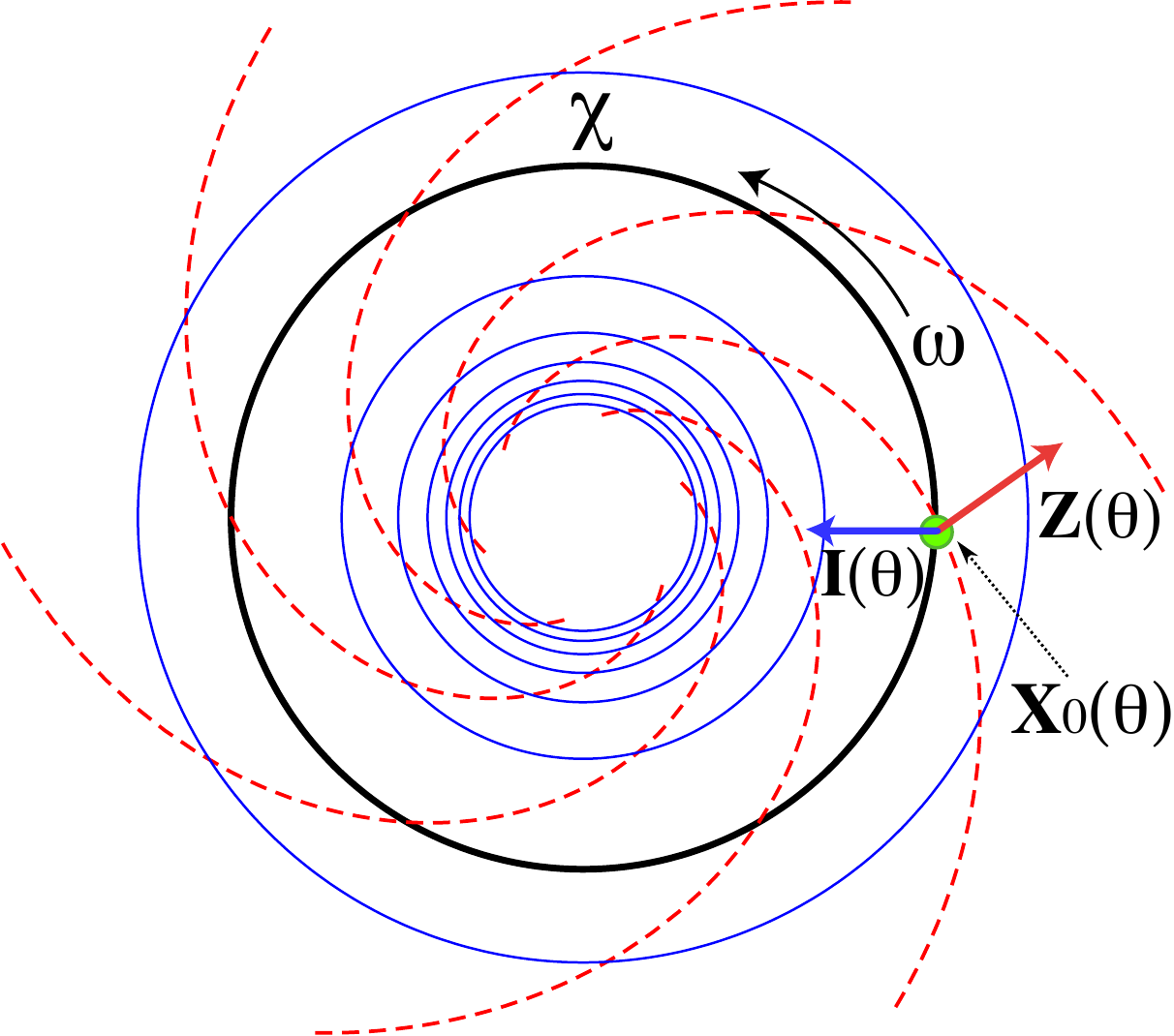}
\hspace{1cm}
%
\caption{Phase and amplitude of a limit cycle. Limit cycle $\chi$ (thick black circle), isochrons (red dashed curves), and isostables (thin blue circles). The green dot shows the oscillator state ${\bf X}_0(\theta)$, red arrow the phase sensitivity function ${\bf Z}(\theta)$, and the blue arrow the amplitude sensitivity function ${\bf I}(\theta)$ at phase $\theta$, respectively.}
\label{fig0}
\end{figure}

In the conventional method of phase reduction, the essential step is the introduction of the {\it asymptotic phase}~\cite{Winfree} (see Fig.~\ref{fig0} for a schematic diagram).
Namely, we assign a scalar phase value $\theta \in [0, 2\pi)$ to the
oscillator
state ${\bf X} \in B$ which eventually converges to the limit cycle $\chi$.
We denote this assignment as $\theta = \Theta({\bf X})$, where $\Theta : B \to [0, 2\pi)$ is a phase function, and require that the phase $\theta$ always increases with a constant frequency $\omega$ as the
oscillator
state ${\bf X}$ evolves according to Eq.~(\ref{eq1}), namely, 
the function
$\Theta$ satisfies
\begin{align}
\dot{\theta} = \dot{\Theta}({\bf X}) = {\nabla_{\bf X}} \Theta({\bf X}) \cdot \dot{\bf X} = {\bf F}({\bf X}) \cdot {\nabla_{\bf X}} \Theta({\bf X}) = \omega
\label{eq2}
\end{align}
for ${\bf X} \in B$, where ${\nabla_{\bf X}} \Theta({\bf X}) \in {\mathbb R}^n$ is the gradient vector of $\Theta$ and $\cdot$ represents the ordinary dot product between two vectors. The level sets of $\Theta$ are called {\it isochrons}.

The phase function $\Theta$ satisfying Eq.~(\ref{eq2}) can be obtained as follows.
For the 
state ${\bf X}_0(t)$ at time $t$ started from a reference state ${\bf X}_R$ at time $0$ on $\chi$, the phase function can be taken as $\Theta({\bf X}_0(t)) = \omega t \ (\mbox{mod}\ 2\pi)$, where ${\bf X}_R$ gives 
the origin of the phase, $\Theta({\bf X}_R) = 0$.
This assigns a phase value $\theta \in [0, 2\pi)$ to 
each point on $\chi$. In what follows, we denote the
state on $\chi$ as ${\bf X}_0(\theta)$ as a function of 
$\theta$.
Note that $\theta = \Theta({\bf X}_0(\theta))$ holds.
To the
state ${\bf X}$ not on $\chi$, we assign a phase value $\Theta({\bf X}) = \theta$ if it converges to the same state on $\chi$ as ${\bf X}_0(\theta)$, namely, if $\lim_{\tau \to \infty} \| S^\tau {\bf X} - S^\tau {\bf X}_0(\theta) \| \to 0,$
where $\| \cdot \|$ is the Euclidean norm and $S^\tau$ represents the time-$\tau$ flow of Eq.~(\ref{eq1}),
satisfying $S^{\tau} {\bf X}(t) = {\bf X}(t+\tau)$.
The phase function $\Theta$ defined as above satisfies Eq.~(\ref{eq2}) for ${\bf X} \in B$.

Next, we consider the amplitude degrees of freedom, representing deviations of the 
state ${\bf X}$ from the limit cycle $\chi$. 
The linear stability of $\chi$ is characterized by the Floquet exponents, $0, \lambda_2, ..., \lambda_n$ in decreasing order of their real parts, where $0$ is associated with the phase direction, namely, the neutral tangential direction along $\chi$, and the real parts of all other exponents $\lambda_2, ..., \lambda_n$ are negative.
For $n$-dimensional oscillators, there are generally $n-1$ amplitudes associated with $\lambda_2, ..., \lambda_n$, but we focus only on the dominant, slowest-decaying amplitude associated with $\lambda_2$ and denote this exponent as $\lambda \ (<0)$, which we assume simple and real for simplicity.

In a similar way to the phase $\theta$, it is useful to assign 
a scalar amplitude $r = R({\bf X})$ to the 
state ${\bf X} \in B$ and assume that $r$ obeys a simple 
equation, where $R : B \to {\mathbb R}$ is an amplitude function.
A natural 
assumption is that $r$ exponentially decays to $0$
as $\dot{r} = \lambda r$ when ${\bf X}$ converges to $\chi$.
Here, the decay rate is given by the Floquet exponent $\lambda$ and $r=0$
when ${\bf X}$ is on $\chi$.
Thus, we require $R$ to satisfy $R({\bf X}_0(\theta)) = 0$ and 
\begin{align}
\dot{r} = \dot{R}({\bf X}) = {\nabla_{\bf X}} R({\bf X}) \cdot \dot{\bf X} = {\bf F}({\bf X}) \cdot {\nabla_{\bf X}} R({\bf X}) = \lambda R({\bf X}) = \lambda r.
\label{eq22}
\end{align}
Recent developments in the Koopman operator approach to nonlinear dynamical systems~\cite{Mauroy,Mauroy2,KoopmanBook} have shown that the above definition is actually natural in the sense that $R$ is given by an eigenfunction of the infinitesimal Koopman operator
${\bf F}({\bf X}) \cdot {\nabla_{\bf X}}$
associated with the eigenvalue $\lambda$, and such $R$ has been 
calculated e.g. for the van der Pol oscillator. The level sets of $R$ are called {\it isostables}, in a similar sense to the isochrons for the asymptotic phase.
In general, we have $n-1$ amplitudes, or principal Koopman eigenfunctions, 
associated
with Floquet exponents $\lambda_2, ..., \lambda_n$, which can take complex values. Accordingly, the range of the 
amplitudes should be taken as ${\mathbb C}$ rather than ${\mathbb R}$.
It is easy to show that the exponential of the phase function, $\Phi = e^{i \Theta}$, is also an eigenfunction of ${\bf F}({\bf X}) \cdot {\nabla_{\bf X}}$ with eigenvalue $i \omega$. Thus, the Koopman operator approach gives a unifying viewpoint on the phase-amplitude description, or global linearization, of the flows around stable limit cycles.

Having defined the phase $\theta = \Theta({\bf X})$ and the amplitude $r = R({\bf X})$ for ${\bf X} \in B$, we can derive approximate phase and amplitude equations for a weakly perturbed limit-cycle oscillator described by
\begin{align}
\dot{\bf X}(t) = {\bf F}({\bf X}(t)) + \epsilon {\bf p}( {\bf X}(t), t),
\label{eq3}
\end{align}
where ${\bf p} \in {\mathbb R}^n$ represents the perturbation applied to the oscillator and 
$\epsilon > 0$ is a small parameter. The equation for the phase
can be expressed as
$
\dot{\theta} = {\nabla_{\bf X}} \Theta({\bf X}) \cdot \dot{\bf X}= {\nabla_{\bf X}} \Theta({\bf X}) \cdot {\bf F}({\bf X}) + \epsilon {\nabla_{\bf X}} \Theta({\bf X}) \cdot {\bf p} = \omega + \epsilon {\nabla_{\bf X}} \Theta({\bf X}) \cdot {\bf p},
$
which still depends on ${\bf X}$. To obtain an equation closed in $\theta$, we use the fact that ${\bf X}$ is 
near ${\bf X}_0(\theta)$ on $\chi$ and approximate ${\bf X}$ as ${\bf X} = {\bf X}_0(\theta) + O(\epsilon)$.
Then, up to the lowest-order
$O(\epsilon)$ in $\epsilon$, the phase $\theta$ obeys
\begin{align}
\dot{\theta}(t) \simeq \omega + \epsilon {\bf Z}(\theta(t)) \cdot {\bf p}({\bf X}_0(\theta(t)), t),
\label{eq5}
\end{align}
where we defined the {\it phase sensitivity function} ${\bf Z}(\theta) = {\nabla_{\bf X}} \Theta({\bf X})|_{{\bf X} = {\bf X}_0(\theta)}$.
Likewise, the amplitude obeys 
$
\dot{r} = {\nabla_{\bf X}} R({\bf X}) \cdot \dot{\bf X}= {\nabla_{\bf X}} R({\bf X}) \cdot {\bf F}({\bf X}) + \epsilon {\nabla_{\bf X}} R({\bf X}) \cdot {\bf p} = \lambda r + \epsilon {\nabla_{\bf X}} R({\bf X}) \cdot {\bf p} 
$
and, by assuming that ${\bf X}$ is in the neighborhood of ${\bf X}_0(\theta)$,
we obtain
\begin{align}
\dot{r}(t)  \simeq \lambda r(t) + \epsilon {\bf I}(\theta(t)) \cdot {\bf p}({\bf X}_0(\theta(t)), t),
\label{eq9}
\end{align}
which is again correct up to $O(\epsilon)$. We here defined ${\bf I}(\theta) = {\nabla_{\bf X}} R({\bf X})|_{{\bf X} = {\bf X}_0(\theta)}$, which we call the {\it amplitude sensitivity function}.

It is difficult to fully determine the phase function $\Theta({\bf X})$ and the amplitude function $R({\bf X})$ for all ${\bf X} \in B$ even numerically in multidimensional systems. However, if we are interested in the weakly perturbed case, Eq.~(\ref{eq3}), we only need the functions ${\bf Z}$ and ${\bf I}$. It can be shown that these quantities are given by $2\pi$-periodic solutions to the 
following
{\it adjoint} linear ODEs
~\cite{Ermentrout1,Kuramoto2,Nakao0,Wilson,Shirasaka,Brown}:
\begin{align}
\omega \frac{d}{d\theta} {\bf Z}(\theta) = - J(\theta)^{\top} {\bf Z}(\theta),
\quad
\omega \frac{d}{d\theta} {\bf I}(\theta) = - [ J(\theta)^{\top} -\lambda ] {\bf I}(\theta),
\label{odeadjoint}
\end{align}
where $J(\theta) = D {\bf F}({\bf X})|_{{\bf X} = {\bf X}_0(\theta)} \in {\mathbb R}^{n \times n}$ is a 
Jacobian matrix of ${\bf F}$
at ${\bf X} = {\bf X}_0(\theta)$ on $\chi$ and $\top$ denotes transpose. The matrix components of $J(\theta)$ are given by $J_{ij}(\theta) = \partial F_i / \partial X_j |_{{\bf X} = {\bf X}_0(\theta)}$ for $i, j = 1, ..., n$, where $F_i$ and $X_i$ are vector components of ${\bf F}$ and ${\bf X}$, respectively.
To be consistent with the definition of the asymptotic phase, ${\bf Z}$ should be normalized as ${\bf Z}(\theta) \cdot {\bf F}({\bf X}_0(\theta)) = \omega$ for $\forall \theta \in [0, 2\pi)$.
The normalization of ${\bf I}$, which determines the scale of $r$, can be chosen arbitrarily.
These adjoint equations can be derived by expanding $\Theta({\bf X})$ and $R({\bf X})$ around ${\bf X}_0(\theta)$ to the first order in ${\bf X} - {\bf X}_0(\theta)$ and plugging them into Eqs.~(\ref{eq2}) and~(\ref{eq22})~\cite{Kuramoto2,Shirasaka,Brown}.

Equations~(\ref{eq5}) and (\ref{eq9}) give the lowest-order phase-amplitude description of the weakly perturbed limit-cycle oscillator Eq.~(\ref{eq3}), from which we can predict the dynamics of the phase $\theta$ and amplitude $r$ and in turn use them to predict the oscillator state near $\chi$.
\footnote{
The system state can be approximately represented as ${\bf X} \simeq {\bf X}_0(\theta) + \epsilon r {\bf u}(\theta)$ up to $O(\epsilon)$, where ${\bf u}(\theta)$ is a $2\pi$-periodic Floquet eigenfunction with Floquet exponent $\lambda$ satisfying $\omega (d/d\theta) {\bf u}(\theta) = [J(\theta) - \lambda ] {\bf u}(\theta)$. See e.g.~\cite{Kuramoto2} for details. This also holds true for the RD system discussed later in Sec.~\ref{sec:3}, namely, ${\bf X}({\bf x}, t)  \simeq {\bf X}_0({\bf x}, \theta) + \epsilon r {\bf u}({\bf x}, \theta)$ where the Floquet eigenfunction ${\bf u}({\bf x}, \theta)$ is $2\pi$-periodic in $\theta$ and
satisfies $\omega (\partial/\partial \theta) {\bf u}({\bf x}, \theta) = [J({\bf x};\theta) - \lambda + D \nabla^2 ] {\bf u}({\bf x}, \theta)$.
}
Note that $\theta$ is decoupled from $r$ at the lowest-order approximation.
Higher-order approximations can also be developed, which yield coupling between $\theta$ and $r$ and more precisely describe the oscillator dynamics~\cite{Wilson2,Kotani2}. In this chapter, we consider the simplest nontrivial lowest-order case and generalize it to PDEs.

\section{Phase-amplitude reduction of reaction-diffusion systems}
\label{sec:3}

We now consider spatially extended reaction-diffusion (RD) systems described by
\begin{align}
\frac{\partial {\bf X}({\bf x}, t)}{\partial t} = {\bf F}({\bf X}({\bf x}, t) ; {\bf x}) + D \nabla^2 {\bf X}({\bf x}, t),
\label{rd1}
\end{align}
where ${\bf X}({\bf x}, t) \in {\mathbb R}^n$ is a $n$-dimensional field variable at position ${\bf x} \in V \subseteq {\mathbb R}^d$ in a $d$-dimensional spatial domain $V$ and at time $t$, representing e.g. concentrations of chemical species involved in the reaction, ${\bf F} : {\mathbb R}^n \times {\mathbb R}^d \to {\mathbb R}^n$ represents position-dependent reaction dynamics of ${\bf X}({\bf x}, t)$, $D \in {\mathbb R}^{n \times n}$ is a matrix of diffusion coefficients, and $\nabla^2$ is a 
Laplacian operator representing the diffusion of ${\bf X}$.
Appropriate boundary conditions, e.g., periodic boundary conditions, are assumed for $V$.
Note that Eq.~(\ref{rd1}) is an infinite-dimensional dynamical system whose system state at $t$ is given by ${\bf X}({\bf x}, t)$ for ${\bf x} \in V$, which we denote as ${\bf X}(\cdot, t) \in C$, where $C$ is some appropriate space of smooth vector-valued functions.
We assume that Eq.~(\ref{rd1}) possesses an exponentially stable limit-cycle solution ${\bf X}_0(\cdot, t)$ of natural period $T$ and frequency $\omega = 2\pi / T$, satisfying ${\bf X}_0({\bf x}, t+T) = {\bf X}_0({\bf x}, t)$ for $\forall {\bf x} \in V$. We again denote this limit-cycle attractor in $C$ as $\chi$ and its basin of attraction as $B \subseteq C$.
Note that this assumption excludes
continuous translational symmetries other than 
the temporal one along the limit cycle. Thus, the system possesses only a single phase variable.

\begin{figure}[t]
\includegraphics[width=0.6\hsize]{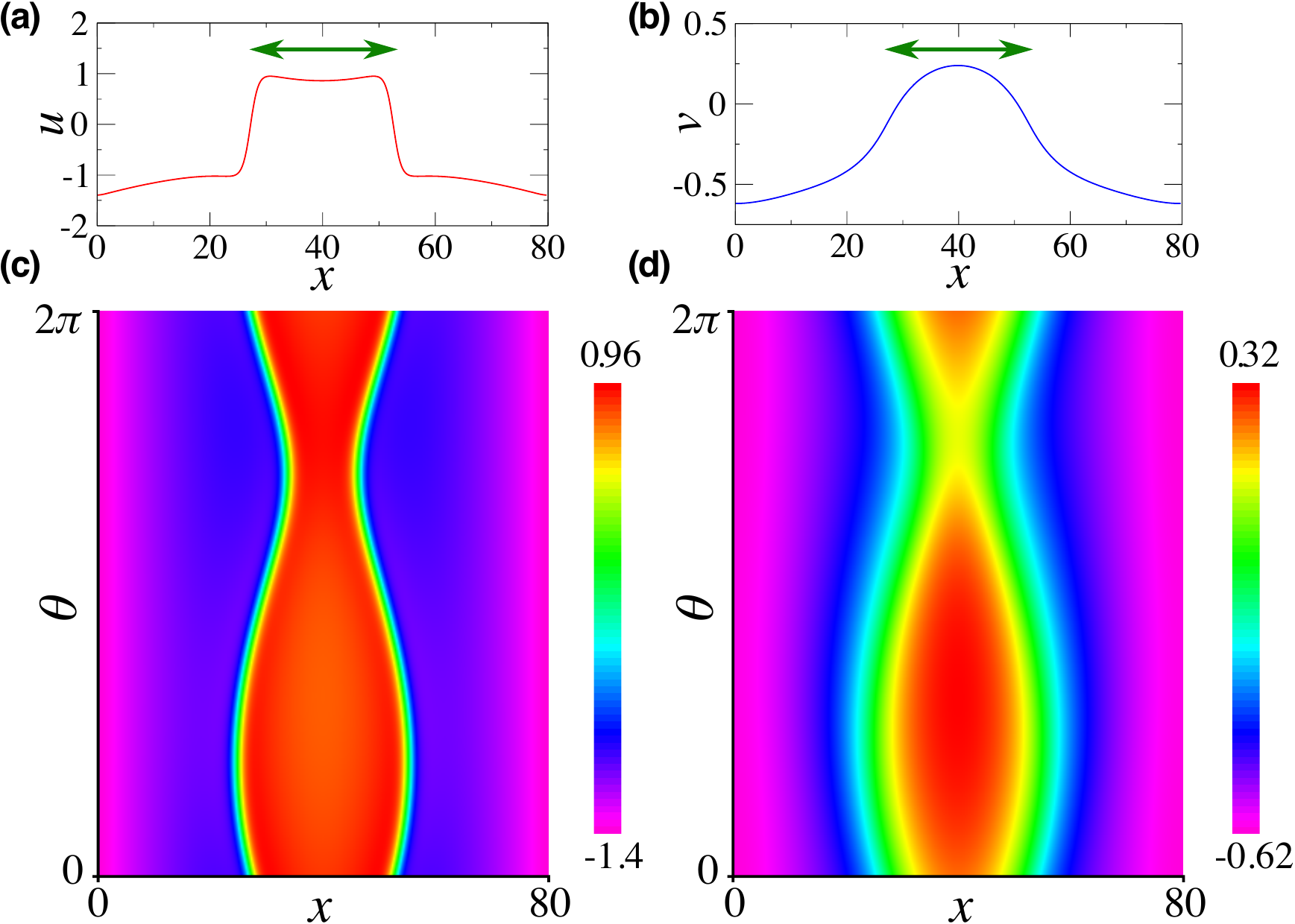}
\hspace{1cm}
\caption{Oscillating-spot solution of the FitzHugh-Nagumo model. (a, b): Spatial profiles of the  $u$ and $v$ components, ${\bf X}(x, 0) = ( u(x, 0), v(x, 0) )^{\top}$, at $\theta = 0$. (c, d): One-period evolution of the $u$ and $v$ components, ${\bf X}(x, \theta) = ( u(x, \theta), v(x, \theta) )^{\top}$, for $0 \leq \theta < 2\pi$.}
\label{fig1}
\end{figure}

Typical examples of limit-cycle solutions of the RD systems are traveling pulses on a ring, oscillating spots, target patterns, and spiral waves in spatially one or two-dimensional systems~\cite{Kuramoto,Nakao}.
Figure~\ref{fig1} shows the oscillating-spot solution of the FitzHugh-Nagumo (FHN) model on a $1$-dimensional interval, where $x \in [0, L]$ with $L=80$ represents the spatial position instead of the vector ${\bf x}$ (see Sec.~\ref{sec5} for details).
Such complex oscillatory patterns in RD systems are difficult to analyze because of their nonlinearity and  infinite-dimensionality.
However, if we are interested in the vicinity of the limit cycle, namely, if we focus on the cases that the oscillatory patterns are only weakly perturbed, we can approximately describe their infinite-dimensional dynamics by simple finite-dimensional phase-amplitude equations in a similar way to the case of ODEs.
In what follows, generalizing the method of phase reduction formulated in Ref.~\cite{Nakao}, we formulate a method of phase-amplitude reduction for RD systems exhibiting stable limit-cycle oscillations.

To this end, we need to introduce the phase $\theta$ and amplitude $r$ of the infinite-dimensional state ${\bf X}(\cdot, t)$ of the RD system. Because we map the field variable to scalars, they should be given by {\it functionals} of ${\bf X}(\cdot, t)$.
We thus define them as $\theta(t) = \Theta[ {\bf X}(\cdot, t) ]$ and $r(t) = R[ {\bf X}(\cdot, t) ]$, where $\Theta : B \to [0, 2\pi)$ is the phase functional and $R : B \to {\mathbb R}$ is the amplitude functional, respectively.
Here, we again focus on the slowest-decaying amplitude associated with the largest non-zero Floquet exponent $\lambda\ (<0)$, which we assume simple, real, and distant from $0$ with a finite spectral gap.

As in the ODE case, we require that these $\theta$ and $r$ obey simple equations, i.e., $\dot \theta = \omega$ and $\dot r = \lambda r$.
Then, from the chain rule of the derivative of functionals, 
\begin{align}
\dot{\theta} &= \dot{\Theta}[ {\bf X}(\cdot, t) ] = 
\int_V \left. \frac{\delta \Theta[{\bf X}(\cdot)]}{\delta {\bf X}({\bf x})} \right|_{{\bf X}({\bf x}) = {\bf X}({\bf x}, t)} \cdot \frac{\partial {\bf X}({\bf x}, t)}{\partial t} d{\bf x} = \omega,
\label{rdphs1}
\\
\dot{r} &= \dot{R}[ {\bf X}(\cdot, t) ] = 
\int_V \left. \frac{\delta R[{\bf X}(\cdot)]}{\delta {\bf X}({\bf x})} \right|_{{\bf X}({\bf x}) = {\bf X}({\bf x}, t)} \cdot \frac{\partial {\bf X}({\bf x}, t)}{\partial t} d{\bf x} = \lambda r
\label{rdamp1}
\end{align}
should hold.
Here, $\partial {\bf X}({\bf x}, t) / \partial t$ is given by Eq.~(\ref{rd1}) and $\delta A[ {\bf X}(\cdot) ] / \delta {\bf X}({\bf x}) |_{{\bf X}({\bf x}) = {\bf X}({\bf x}, t)}\in {\mathbb R}^n$  ($A = \Theta, R$) represents the functional derivative of the functional $A[ {\bf X} ]$ with respect to ${\bf X}({\bf x})$ evaluated at ${\bf X}({\bf x}) = {\bf X}({\bf x}, t)$.
Such phase and amplitude functionals can, in principle, be defined 
as in the ODE case.
We denote the system state on $\chi$ as ${\bf X}_0(\cdot, \theta)$ as a function of the phase $\theta \in [0,2\pi)$.
Note that $\Theta[{\bf X}_0(\cdot, \theta)] = \theta$ and $R[{\bf X}_0(\cdot, \theta)] = 0$ hold.

Thus, Eqs.~(\ref{eq2}) and (\ref{eq22}) for ODEs can formally be generalized to spatially extended RD systems, where the vector gradient is replaced by the functional derivative.
We can also interpret $R$ and $\Phi = e^{i\Theta}$ as eigenfunctionals of the infinitesimal Koopman operator for RD systems~\cite{Nakao2,Nakao3}.
We note here that the explicit forms of the functionals $\Theta$ and $R$, which are difficult to obtain, are not necessary in the following derivation of the phase and amplitude equations near the limit-cycle solution.
\begin{figure}[tb]
\includegraphics[width=0.6\hsize]{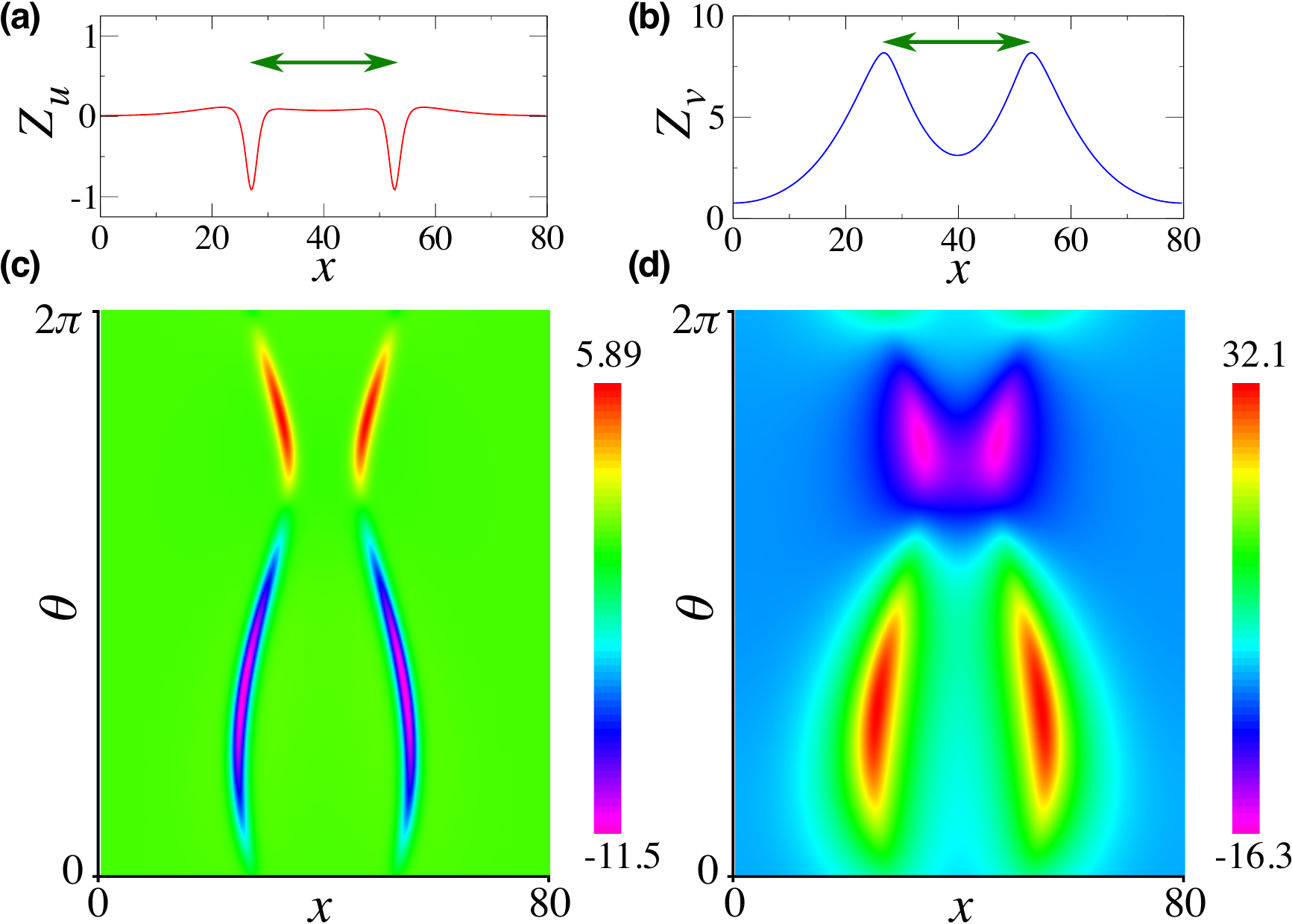}
\hspace{1cm}
\caption{Phase sensitivity function of the oscillating-spot solution of the FitzHugh-Nagumo model. (a, b): Spatial profiles of the $u$ and $v$ components, ${\bf Z}(x, 0) = ( Z_u(x, 0), Z_v(x, 0) )^{\top}$, at $\theta = 0$. (c, d): One-period evolution of the $u$ and $v$ components, ${\bf Z}(x, \theta) = ( Z_u(x, \theta), Z_v(x, \theta) )^{\top}$, for $0 \leq \theta < 2\pi$.}
\label{fig2}
\end{figure}

Let us now consider the case that the RD system Eq.~(\ref{rd1}) is weakly perturbed as
\begin{align}
\frac{\partial {\bf X}({\bf x}, t)}{\partial t} = {\bf F}({\bf X}({\bf x}, t) ; {\bf x}) + D \nabla^2 {\bf X}({\bf x}, t) + \epsilon {\bf p}({\bf X}(\cdot, t), {\bf x}, t),
\label{rd2}
\end{align}
where 
$\epsilon > 0$ is a small parameter and ${\bf p}({\bf X}(\cdot, t), {\bf x}, t) \in {\mathbb R}^n$ is the applied perturbation that can depend on the state ${\bf X}$, position ${\bf x}$, and time $t$.
In a similar way to the case of ODEs, using the phase and amplitude functionals $\Theta$ and $R$ satisfying Eqs.~(\ref{rdphs1}) and (\ref{rdamp1}), we can reduce Eq.~(\ref{rd2}) to a set of approximate phase-amplitude equations as
\begin{align}
\dot{\theta} &= 
\int_V \left. \frac{\delta \Theta[{\bf X}(\cdot)]}{\delta {\bf X}({\bf x})} \right|_{{\bf X}({\bf x}) = {\bf X}({\bf x}, t)} \cdot \frac{\partial {\bf X}({\bf x}, t)}{\partial t} d{\bf x} 
\simeq
\omega + \epsilon \int_V {\bf Z}({\bf x}, \theta) \cdot {\bf p}({\bf X}_0(\cdot, \theta), {\bf x}, t) d{\bf x},
\label{rdphs2}
\\
\dot{r} &=
\int_V \left. \frac{\delta R[{\bf X}(\cdot)]}{\delta {\bf X}({\bf x})} \right|_{{\bf X}({\bf x}) = {\bf X}({\bf x}, t)} \cdot \frac{\partial {\bf X}({\bf x}, t)}{\partial t} d{\bf x} 
\simeq
\lambda r + \epsilon \int_V {\bf I}({\bf x}, \theta) \cdot {\bf p}({\bf X}_0(\cdot, \theta), {\bf x}, t) d{\bf x},
\label{rdamp2}
\end{align}
where $\partial {\bf X}({\bf x}, t) / \partial t$ is given by Eq.~(\ref{rd2}). 
Here, in the last expression in each equation,
we have approximately evaluated the functional derivatives 
and perturbations in the integral at ${\bf X}_0({\bf x}, \theta)$ rather than at ${\bf X}({\bf x}, t)$,
assuming that the system state is sufficiently close to $\chi$,
and introduced the phase and amplitude sensitivity functions 
\begin{align}
{\bf Z}({\bf x}, \theta) = \left. \frac{\delta \Theta[{\bf X}(\cdot)]}{\delta {\bf X}({\bf x})} \right|_{{\bf X}({\bf x}) = {\bf X}_0({\bf x}, \theta)},
\quad
{\bf I}({\bf x}, \theta) = \left. \frac{\delta R[{\bf X}(\cdot)]}{\delta {\bf X}({\bf x})} \right|_{{\bf X}({\bf x}) = {\bf X}_0({\bf x}, \theta)},
\end{align}
which now depend also on the position ${\bf x}$. The approximate equations~(\ref{rdphs2}) and (\ref{rdamp2}) are correct up to $O(\epsilon)$ like Eqs.~(\ref{eq5}) and (\ref{eq9}) in the ODE case.

\begin{figure}[tb]
\includegraphics[width=0.6\hsize]{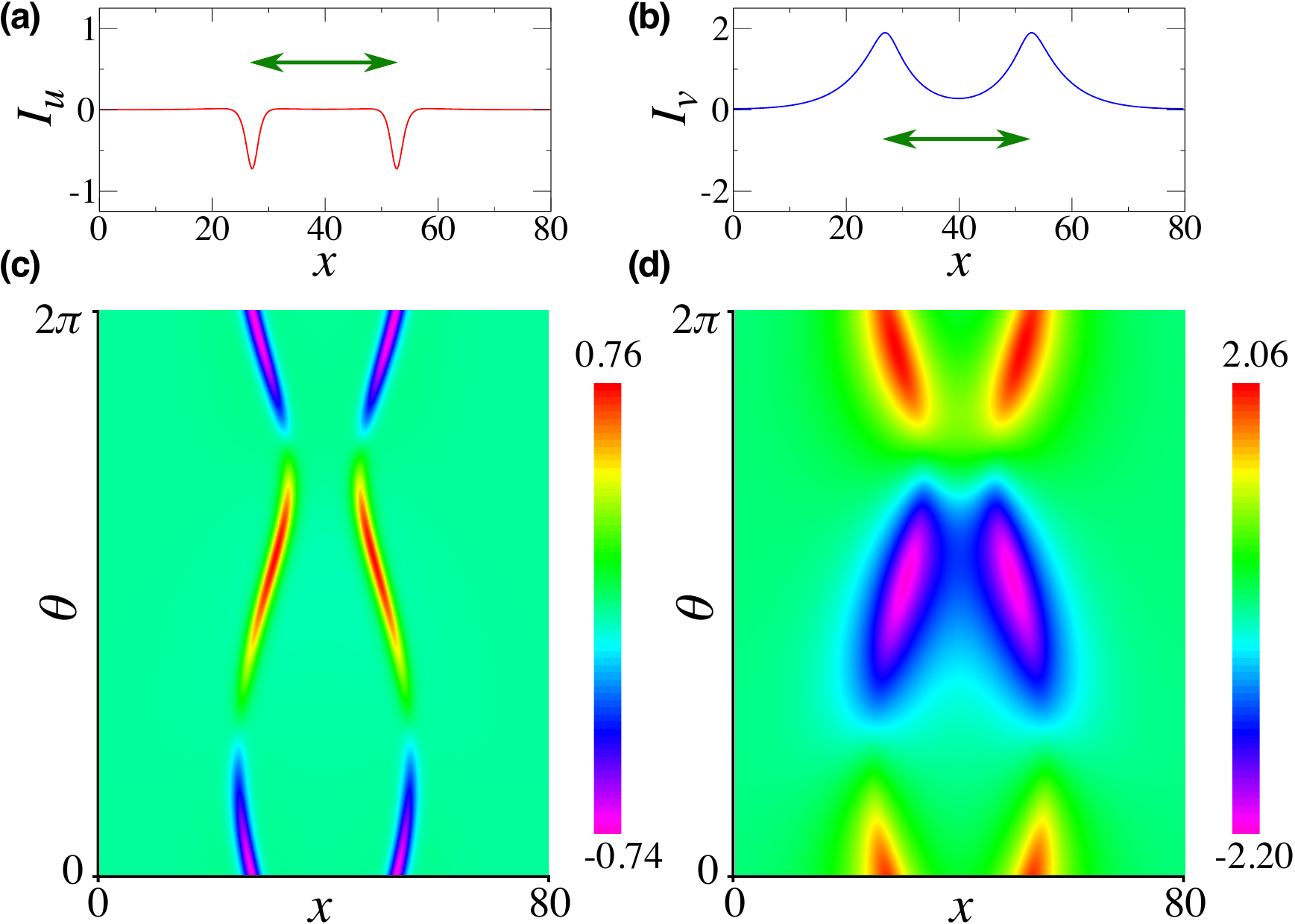}
\hspace{1cm}
\caption{Amplitude sensitivity function of the oscillating-spot solution of the FitzHugh-Nagumo model. (a, b): Spatial profiles of the $u$ and $v$ components, ${\bf I}(x, 0) = ( I_u(x, 0), I_v(x, 0) )^{\top}$, at $\theta = 0$. (c, d): One-period evolution of the $u$ and $v$ components, ${\bf I}(x, \theta) = ( I_u(x, \theta), I_v(x, \theta) )^{\top}$, for $0 \leq \theta < 2\pi$.}
\label{fig3}
\end{figure}

Though it is difficult to obtain the functionals $\Theta$ and $R$ explicitly, it can be shown that the sensitivity functions ${\bf Z}$ and ${\bf I}$ are given by $2\pi$-periodic solutions to the following adjoint linear PDEs~\cite{Nakao,Nakao2}:
\begin{align}
\omega \frac{\partial {\bf Z}({\bf x}, \theta)}{\partial \theta} &= - J({\bf x};\theta)^{\top} {\bf Z}({\bf x}, \theta) - D^{\top} \nabla^2 {\bf Z}({\bf x}, \theta),
\cr
\omega \frac{\partial {\bf I}({\bf x}, \theta)}{\partial \theta} &= - [ J({\bf x};\theta)^{\top} - \lambda ] {\bf I}({\bf x}, \theta) - D^{\top} \nabla^2 {\bf I}({\bf x}, \theta),
\label{adjpde}
\end{align}
where $J({\bf x};\theta) = D{\bf F}({\bf X}({\bf x}, t) ; {\bf x}) \in {\mathbb R}^{n \times n}$ is the Jacobian matrix of ${\bf F}$ evaluated at position ${\bf x}$.
The normalization for ${\bf Z}$ is now given by $\int_V {\bf Z}({\bf x}, \theta) \cdot \{ {\bf F}({\bf X}_0({\bf x}, \theta)) + D \nabla^2 {\bf X}_0({\bf x}, \theta) \} d{\bf x} = \omega$ for $\forall \theta \in [0,2\pi)$.
These equations are straightforward generalization of the adjoint equations~(\ref{odeadjoint}) and the normalization condition for the ODE case.
Figures~\ref{fig2} and~\ref{fig3} show the phase and amplitude sensitivity functions of the oscillating-spot solution of the FHN model, respectively (see Sec.~\ref{sec5} for details).

Thus, by defining the phase and amplitude functionals, we can reduce the weakly perturbed RD system, Eq.~(\ref{rd2}), to a set of ODEs~(\ref{rdphs2}) and (\ref{rdamp2}) for the phase and amplitude.
The reduced phase-amplitude equations are much simpler than the original RD system and facilitate detailed analysis of the oscillatory patterns. 
In Ref.~\cite{Nakao}, the phase equation has been used to analyze synchronization between a pair of mutually coupled RD systems.
In the next section, we analyze optimal entrainment with feedback stabilization of RD systems using the phase and amplitude equations.

\section{Optimal entrainment of oscillatory patterns with feedback}

As an application of the reduced phase-amplitude equations, we analyze entrainment of a RD system exhibiting oscillatory patterns by an optimized periodic forcing, generalizing Ref.~\cite{Zlotnik} for limit-cycle oscillators described by ODEs. The model is
\begin{align}
\frac{\partial {\bf X}({\bf x}, t)}{\partial t} = {\bf F}({\bf X}({\bf x}, t) ; {\bf x}) + D \nabla^2 {\bf X}({\bf x}, t) + \epsilon K {\bf q}({\bf x}, \Omega t),
\label{rd3}
\end{align}
where ${\bf q}({\bf x}, \Omega t) \in {\mathbb R}^n$ represents a temporally periodic smooth forcing pattern of frequency $\Omega$ satisfying ${\bf q}({\bf x}, \Omega t + 2\pi) = {\bf q}({\bf x}, \Omega t)$ and $K = \mbox{diag} \{K_1, \cdots, K_n \}  \in {\mathbb R}^{n \times n}$ is a constant diagonal matrix representing the effect of ${\bf q}$ on ${\bf X}$ (e.g., $K_{1} \neq 0$ and $K_{2, ..., n} = 0$ when only the 1st component of ${\bf X}$ is driven by the forcing).
We assume that the forcing frequency $\Omega$ is sufficiently close to the natural frequency $\omega$ of the system, namely, the frequency mismatch $\omega - \Omega$ is a small value of $O(\epsilon)$ and denote it as $\epsilon \Delta$ where $\Delta = O(1)$.
The reduced approximate phase-amplitude equations are
\begin{align}
\dot{\theta}(t) &= \omega + \epsilon \int_V {\bf Z}({\bf x}, \theta(t)) \cdot K {\bf q}({\bf x}, \Omega t) d{\bf x},
\cr
\dot{r}(t) &= \lambda r(t) + \epsilon \int_V {\bf I}({\bf x}, \theta(t)) \cdot K {\bf q}({\bf x}, \Omega t) d{\bf x}.
\end{align}

For the 
linear stability analysis of the entrained state, we only need the phase equation. 
Following the standard procedure~\cite{Kuramoto}, we consider the phase difference $\phi = \theta - \Omega t$ between the system and periodic forcing, which obeys
\begin{align}
\dot{\phi}(t) = \epsilon \Delta + \epsilon \int_V {\bf Z}({\bf x}, \phi(t)+\Omega t) \cdot K {\bf q}({\bf x}, \Omega t) d{\bf x}.
\end{align}
Because the right-hand side is $O(\epsilon)$, $\phi$ is a slowly varying quantity and the right-hand side can be averaged over one period of the forcing with fixed $\phi$. We thus obtain
\begin{align}
\dot{\phi}(t) = \epsilon [ \Delta + \Gamma(\phi(t)) ],
\label{phsdif}
\end{align}
which is correct up to $O(\epsilon)$, where we defined a $2\pi$-periodic {\it phase coupling function}
\begin{align}
\Gamma(\phi)  
= \frac{1}{2\pi} \int_0^{2\pi} 
\int_V {\bf Z}({\bf x}, \phi + \psi) \cdot K {\bf q}({\bf x}, \psi) d{\bf x} 
d\psi 
= \langle {\bf Z}({\bf x}, \phi + \psi) \cdot K {\bf q}({\bf x}, \psi)  \rangle.
\end{align}
Here, we introduced the abbreviation $\langle A({\bf x}, \psi) \rangle = (2\pi)^{-1} \int_0^{2\pi} \left\{ \int_V  A({\bf x}, \psi) d{\bf x} \right\} d\psi$.
Equation~(\ref{phsdif}) can possess a stable fixed point $\phi^* \in [0, 2\pi)$ satisfying $\Delta + \Gamma(\phi^*) = 0$ and $\Gamma'(\phi^*) < 0$ when $\Delta$ is in an appropriate range, whose linear stability is given by the slope $\Gamma'(\phi^*) = d\Gamma(\phi)/d\phi|_{\phi = \phi^*}$ of $\Gamma(\phi)$ at $\phi^*$.
The oscillatory pattern 
can be entrained to the periodic forcing when such a stable fixed point $\phi^*$ exists. 

\begin{figure}[tb]
\includegraphics[width=0.85\hsize]{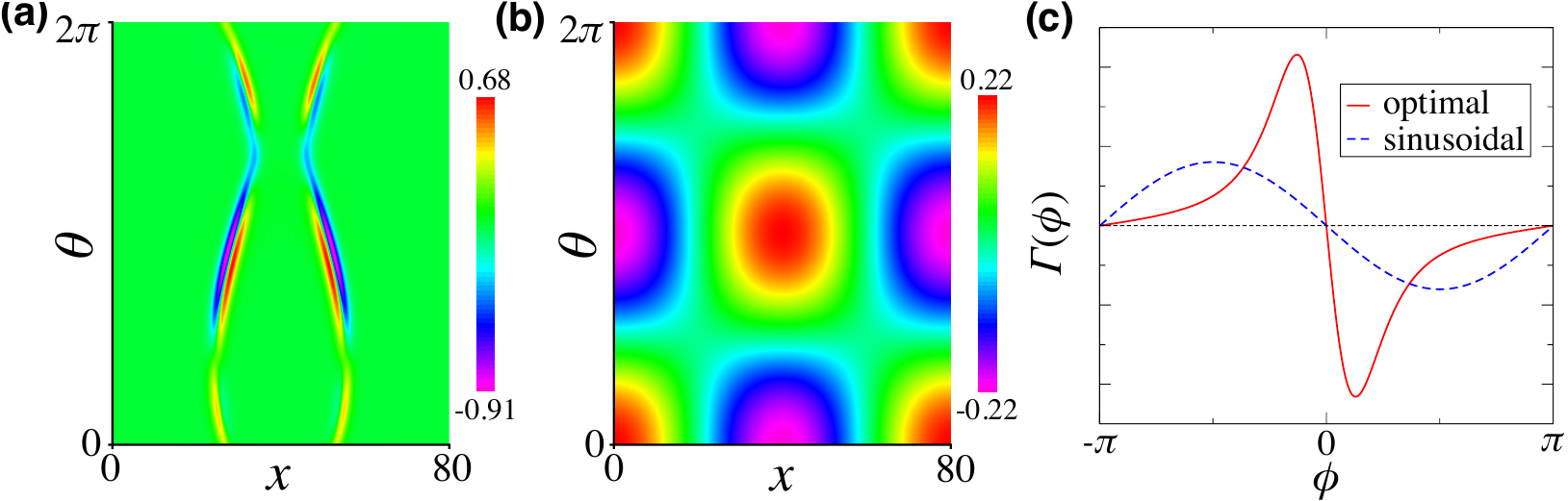}
\caption{Forcing patterns and phase coupling functions. (a, b): One-period evolution of the $u$ component of the (a) optimal forcing ${\bf q}_{\mbox{opt}}(x, \theta)$ and (b) sinusoidal forcing ${\bf q}_{\mbox{sin}}(x, \theta)$ for $0 \leq \theta < 2\pi$. (c): Phase coupling functions $\Gamma(\phi)$ for the optimal and sinusoidal forcing patterns.}
\label{fig4}
\end{figure}

We seek the optimal periodic forcing ${\bf q}({\bf x}, \psi)$ for stable entrainment, which minimizes $\Gamma'(\phi^*)$ under the constraint that the power of ${\bf q}({\bf x}, \psi)$ is fixed at $P > 0$,
i.e.,
$
\langle \| {\bf q}({\bf x}, \psi) \|^2 \rangle = P,
$
and also under the constraint that Eq.~(\ref{phsdif}) has a fixed point at given $\phi^*$, i.e., $\Delta + \Gamma(\phi^*) = 0$ holds. Thus, we solve an optimization problem
\begin{align}
\mbox{maximize} \ -\Gamma'(\phi^*) \quad \mbox{subject to} \quad \langle \| {\bf q}({\bf x}, \psi) \|^2 \rangle = P, \  \Delta + \Gamma(\phi^*) = 0.
\end{align}
To this end, we define an objective functional,
\begin{align}
S[{\bf q}(\cdot, \psi)] = - \Gamma'(\phi^*) + \zeta  \left\{ \langle \| {\bf q}({\bf x}, \psi) \|^2 \rangle - P \right\} + \mu \{ \Delta + \Gamma(\phi^*) \},
\end{align}
where $\zeta$ and $\mu$ are Lagrange multipliers.
Solving the stationarity condition, $\delta S / \delta {\bf q}({\bf x}, \psi) = 0$, and eliminating $\mu$ by using the second constraint as well as the $2\pi$-periodicity of ${\bf Z}({\bf x}, \theta)$ in $\theta$, the optimal periodic forcing can be obtained as
\begin{align}
{\bf q}_{\mbox{\small opt}}({\bf x}, \psi) = \frac{1}{2 \zeta } K^{\top} \partial_\psi {\bf Z}(\phi^* + \psi) - \frac{ \Delta }{ \langle  \| K^{\top} {\bf Z}({\bf x}, \phi^* + \psi) \|^2 \rangle } K^{\top} {\bf Z}(\phi^* + \psi)
\label{optforce}
\end{align}
and the optimized stability exponent as
\begin{align}
\Gamma'_{\mbox{\small opt}}(\phi^*) = \frac{1}{2 \zeta } \langle \| K^{\top} \partial_\psi {\bf Z}(\phi^*+\psi) \|^2 \rangle,
\end{align}
where the multiplier $\zeta $ is calculated from the first constraint as
\begin{align}
\zeta  = - \frac{1}{2} \left( 
\frac{ \langle \| K^{\top} \partial_\psi {\bf Z}({\bf x}, \phi^*+\psi) \|^2 \rangle }{P - { \Delta^2} / {\langle \| K^{\top} {\bf Z}({\bf x}, \phi^*+\psi)\|^2 \rangle } } \right)^{1/2}.
\end{align}
Figure~\ref{fig4} shows the optimized forcing, together with a spatiotemporally sinusoidal forcing for comparison, and the resulting phase coupling functions for the oscillating-spot solution of the FHN model (see Sec.~\ref{sec5} for details).

As we demonstrate in the next section, the optimized forcing ${\bf q}_{\mbox{\small opt}}$ gives higher stability of the entrained state than the sinusoidal forcing pattern ${\bf q}_{\mbox{\small sin}}$ of the same power.
However, it can also happen that the optimized pattern perturbs the system too efficiently and kicks the system state far away from the limit cycle, where the reduced equations are no longer accurate.
If so, we may need to decrease the forcing power and may not be able to improve the stability of the entrainment as desired.

In order to cope with this problem, we consider a simple feedback stabilization of the oscillatory pattern, which suppresses the deviation of the system state from the unperturbed limit cycle $\chi$.
To this end, we evaluate the phase $\theta = \Theta[{\bf X}(\cdot, t)]$ of the system state ${\bf X}(\cdot, t)$, calculate the difference vector
$
{\bf y}({\bf x}, t) = {\bf X}({\bf x}, t) - {\bf X}_0({\bf x}, \theta),
$
and apply a feedback forcing of the form $- \alpha {\bf y}({\bf x}, t)$ to the RD system, Eq.~(\ref{rd3}), as
\begin{align}
\frac{\partial {\bf X}({\bf x}, t)}{\partial t} = {\bf F}({\bf X}({\bf x}, t) ; {\bf x}) + D \nabla^2 {\bf X}({\bf x}, t) + \epsilon K {\bf q}({\bf x}, \Omega t) - \epsilon \alpha {\bf y}({\bf x}, t),
\label{rd4}
\end{align}
where $\alpha > 0$ is the feedback gain~\footnote{We here simply assume that the whole spatial pattern can be directly observed. This may not be realistic in practical control problems and some approximate methods may have to be devised.}.

When ${\bf X}$ is sufficiently close to $\chi$, we can show that this ${\bf y}$ is (bi-)orthogonal to the phase sensitivity function ${\bf Z}$. Indeed, we can express the phase of ${\bf X}$ as
\begin{align}
\theta = \Theta[ {\bf X}(\cdot, t) ] = \Theta[ {\bf X}_0(\cdot, \theta) + {\bf y}({\bf x}, t) ] \simeq
\Theta[ {\bf X}_0(\cdot, \theta) ] + \int_V {\bf Z}({\bf x}, \theta) \cdot {\bf y}({\bf x}, t) d{\bf x} 
\end{align}
by retaining the lowest-order functional Taylor expansion of $\Theta$ in ${\bf y}$. Therefore, $\int_V {\bf Z}({\bf x}, \theta) \cdot {\bf y}(t) d{\bf x} = 0$ holds at the lowest order because $\theta = \Theta[{\bf X}_0(\cdot, \theta)]$.
Thus, the reduced phase equation of Eq.~(\ref{rd4}) is the same as that for Eq.~(\ref{rd3}) and the feedback forcing term $- \epsilon \alpha {\bf y}({\bf x}, t)$ does not affect the phase dynamics of the system at the lowest order.
On the other hand, the amplitude $r$ of ${\bf X}(\cdot, t)$ is expressed as
\begin{align}
r = R[{\bf X}(\cdot, t)] = R[ {\bf X}_0(\cdot, \theta) + {\bf y}({\bf x}, t) ] \simeq \int_V {\bf I}({\bf x}, \theta) \cdot {\bf y}({\bf x}, t) d{\bf x}
\label{eq28}
\end{align}
where $R[{\bf X}_0(\cdot, \theta)] = 0$ by definition. Therefore, at the lowest order, the amplitude equation for Eq.~(\ref{rd4}) is given by
\begin{align}
\dot{r}(t)
= ( \lambda - \epsilon \alpha ) r(t) + \int_V {\bf I}({\bf x}, \theta(t)) \cdot K {\bf q}({\bf x}, \Omega t) d{\bf x}.
\end{align}
Thus, by introducing the feedback term $- \alpha {\bf y}({\bf x}, t)$, we can improve the linear stability of $\chi$ from $\lambda$ to $\lambda - \epsilon \alpha$ and keep the system state closer to $\chi$ while not affecting the phase dynamics, which allows us to apply periodic forcing with larger power.

\section{Example: Oscillating spot in the FitzHugh-Nagumo model}
\label{sec5}

As an example of the RD system possessing a stable limit cycle, we consider the FitzHugh-Nagumo model in one dimension, which exhibits a localized oscillating-spot pattern (see Ref.~\cite{Nakao} for details). The model is given by Eq.~(\ref{rd3}) with 
\begin{align}
{\bf X}(x, t) = \begin{pmatrix} u \\ v \end{pmatrix}, 
\quad 
{\bf F}({\bf X}; x) = \begin{pmatrix} u ( u - \beta(x)  ) ( 1 - u ) - v \\ \epsilon (u - \gamma v) \end{pmatrix},
\quad
D = \begin{pmatrix} \kappa & 0 \\ 0 & \delta \end{pmatrix},
\end{align}
where $u(x, t)$ and $v(x, t)$ are the activator and inhibitor field variables at time $t$ and position $x$ ($0 \leq x \leq L$), respectively, $\beta(x) $, $\epsilon$, and $\tau$ are parameters, and $\kappa$ and $\delta$ are diffusion coefficients of $u$ and $v$.
We consider a system of length $L=80$ and assume no-flux boundary conditions $\partial {\bf X}(0,t) / \partial x = \partial {\bf X}(L, t) / \partial x = 0$ at $x=0$ and $L$.
In order to pin the spot to the center $x=L/2$ of the system, we assume that $\beta(x) $ is position-dependent and is given by $\beta(x) = \beta_0 + ( \beta_1 - \beta_0 ) ( x/L - 1/2 )^2$ with $\beta_0 = -1.1$ and $\beta_1 = -1.6$.
The other parameters are $\gamma = 2$ and $\epsilon = 0.0295$, and the diffusion coefficients are $\kappa = 1$ and $\delta = 2.5$.
By choosing an appropriate initial condition, this system converges to a stable limit cycle $\chi$ :  
 ${\bf X}_0(x, \theta) = ( u_0(x, \theta), v_0(x, \theta) )^{\top}$ ($0 \leq x \leq L$, $0 \leq \theta < 2\pi$) with natural period $T \simeq 196.5$ and frequency $\omega \simeq 0.032$ corresponding to the oscillating spot.
The second Floquet exponent of $\chi$ is real and evaluated as $\lambda \simeq -0.387$.
Using the adjoint equations~(\ref{adjpde}), we can calculate the phase and amplitude sensitivity functions ${\bf Z}$ and ${\bf I}$ of $\chi$.

\begin{figure}[tb]
\includegraphics[width=0.7\hsize]{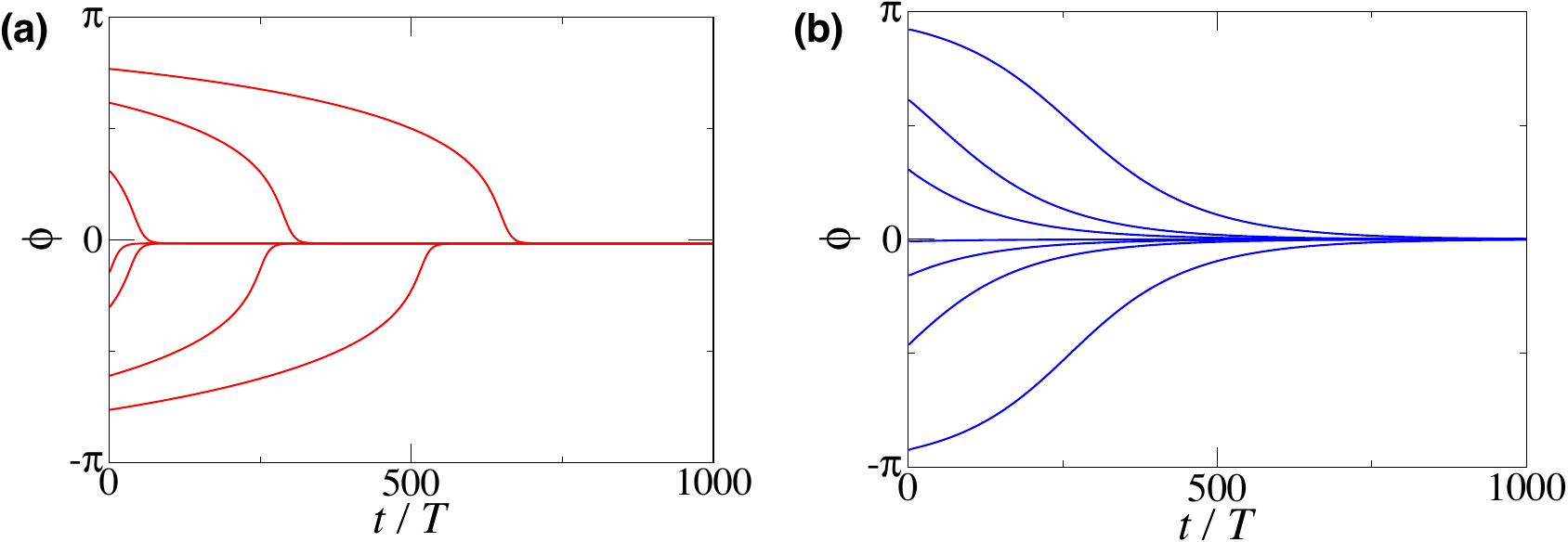}
\caption{Entrainment by the (a) optimized and (b) sinusoidal forcing. Evolution of the phase difference $\phi = \theta - \Omega t$ between the system and the periodic forcing, started from different initial conditions.
}
\label{fig5}
\end{figure}

Figure~\ref{fig1} shows the snapshot and one period evolution of the limit-cycle solution ${\bf X}_0$, and Figs.~\ref{fig2} and~\ref{fig3} show the snapshot and one-period evolution of the phase and amplitude sensitivity functions ${\bf Z}$ and ${\bf I}$, respectively.
It can be seen that both sensitivity functions take large values near the domain walls of the oscillating spot, indicating that perturbations given to the domain walls have strong influence on the phase and amplitude of the system.  Reflecting the difference in the diffusion coefficients, the patterns of the $u$ component are sharper than those of $v$.

We consider entrainment of this oscillating-spot solution by the optimized periodic forcing. For simplicity, we assume that the forcing frequency $\Omega$ is equal to the natural frequency $\omega$ of the system, i.e., $\Delta = 0$. In this case, we can set the value of $\phi^*$ arbitrarily by shifting the origin of the phase, and we fix it as $\phi^* = 0$.
We set $\epsilon = 0.01$ and $K = \mbox{diag\ } \{K_1, K_2\} = \mbox{diag\ } \{0.05, 0\}$,
namely, we apply the periodic forcing only to the $u$ component of the system.
We calculate the optimal forcing ${\bf q}_{\mbox{\small opt}}(x, \psi) = (q_{\mbox{\small opt}}^u(x, \psi), 0)^{\top}$ of power $P=1$ by Eq.~(\ref{optforce}), and also a spatiotemporally sinusoidal forcing ${\bf q}_{\mbox{\small sin}}(x, \psi) \propto ( \cos ( 2 \pi x / L ) \sin \psi, 0 )^{\top}$ of the same power, and drive the system periodically with these forcing patterns.

Figures~\ref{fig4}(a) and (b) show the $u$ component of the forcing patterns ${\bf q}_{\mbox{\small opt}}(x, \psi)$ and ${\bf q}_{\mbox{\small sin}}(x, \psi)$ for one period of oscillation, respectively, and Fig.~\ref{fig4}(c) shows the resulting phase coupling functions $\Gamma(\phi)$. It can be seen that the optimal forcing pattern selectively perturbs the domain walls where the system's phase sensitivity is high, and the slope $\Gamma'(0)$ determining the linear stability of the entrained state $\phi^*=0$ is much larger in the optimized case than in the sinusoidal case.

Figure~\ref{fig5} shows the actual convergence of the phase difference $\phi = \theta - \Omega t$ between the system and periodic forcing to $\phi^*=0$ from several different initial conditions, obtained by direct numerical simulations of Eq.~(\ref{rd3}). It can be confirmed that the asymptotic convergence to the fixed point at $\phi^*=0$ is faster and the entrainment is established earlier in the optimized case.

Now, as we discussed in the previous section, when the periodic forcing is not sufficiently small, it may kick the system state far away from the unperturbed limit cycle and can lead to breakdown of the lowest-order phase-amplitude description.
By introducing the feedback stabilization, we may be able to keep the system state close to the limit cycle and apply stronger forcing.
In order to confirm this, we numerically simulate Eq.~(\ref{rd4}) with the feedback forcing term. In the numerical calculation, the phase $\theta$ of the system state ${\bf X}$ is evaluated by using the procedure described below Eq.~(\ref{eq2}) in Sec.~2, which also applies to the RD case, and then the amplitude $r$ of ${\bf X}$ is evaluated using Eq.~(\ref{eq28}) with ${\bf y}({\bf x}, t) = {\bf X}({\bf x}, t) - {\bf X}_0({\bf x}, \theta)$ within linear approximation (note that ${\bf X}$ is in the neighborhood of ${\bf X}_0(\cdot, \theta)$ when the perturbation is weak).

\begin{figure}[tb]
\includegraphics[width=0.75\hsize]{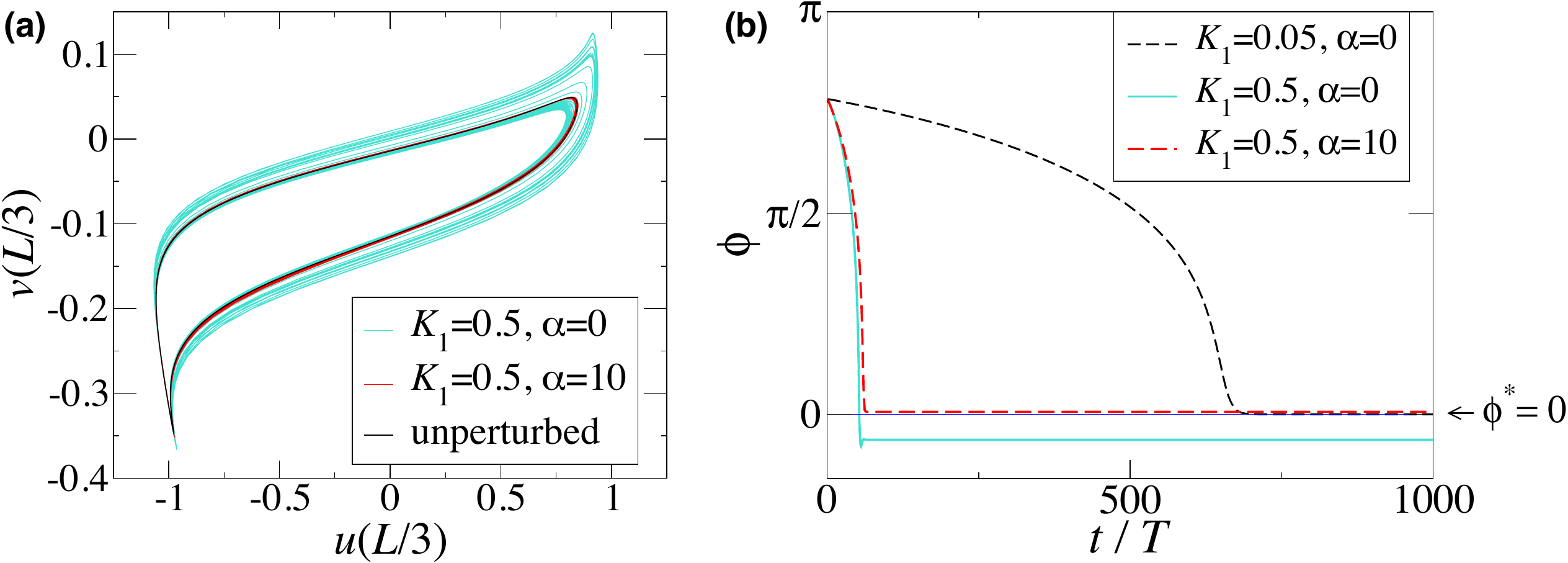}
\caption{Effect of strong forcing and feedback. (a): Trajectories of the system state $(u(x, t), v(x, t))^{\top}$ at $x=L/3$ for the cases with (i) no forcing, (ii) strong forcing without feedback ($K_1=0.5$, $\alpha = 0$), and (iii) strong forcing with feedback ($K_1=0.5$, $\alpha=10$). (b): Convergence of the phase difference $\phi = \theta - \Omega t$ for the case only with weak forcing, $K_1=0.05$, and for the cases (ii) and (iii).}
\label{fig6}
\end{figure}

Figure~\ref{fig6}(a) shows the trajectory of the field variable $(u, v)^{\top}$ at $x=L/3$ obtained by direct numerical simulations of Eq.~(\ref{rd4}) for the cases with (i) no forcing (black), (ii) strong forcing $K_1=0.5$ without feedback, $\alpha=0$ (light blue), and (iii) strong forcing $K_1=0.5$ with feedback gain $\alpha = 10$ (red, overlaps with the black curve).
The curve for the case (i) corresponds to the unperturbed limit cycle $\chi$. We can observe that the strong forcing without feedback drives the system state away from $\chi$ in the case (ii), but the introduction of the feedback keeps the system state close to $\chi$ in the case (iii).
Figure~\ref{fig6}(b) shows the convergence of the phase difference $\phi$
for the cases (ii) and (iii). For comparison, the evolution of $\phi$ under weak forcing $K_1=0.05$ without feedback is also shown.
In the case (ii) without feedback, the final phase difference is considerably different from the target value $\phi^* = 0$ because of the large deviation of the system state from $\chi$.\footnote{In this case, the system state converges to a spurious periodic orbit after initial transient, which is induced by the effect of the strong periodic forcing and larger than the original limit cycle $\chi$.} In contrast, in the case (iii) with feedback, $\phi$ converges to the correct target value $\phi^*=0$ even though the forcing is 10 times stronger.
Thus, the feedback stabilization allows us to apply much stronger periodic forcing and realize faster entrainment.

\section{Summary}
\label{sec:4}

We have formulated the method of phase and amplitude description for limit-cycle oscillations in spatially extended RD systems subjected to weak perturbations.
Though the RD systems are infinite-dimensional dynamical systems, we can still approximately describe them by a set of phase and amplitude equations in the vicinity of the unperturbed limit cycle, which can be used for the analysis and control of complex oscillatory dynamics in RD systems.
As an application, we have considered entrainment of the oscillatory pattern with optimized periodic forcing and feedback stabilization, which allows us to apply stronger forcing while keeping the approximate phase-amplitude description valid and thereby realizing more stable entrainment.

The method of phase reduction for infinite-dimensional dynamical systems has also been developed for delay-differential equations exhibiting limit-cycle oscillations~\cite{Kotani,Novicenko}, for nonlinear Fokker-Planck (integro-differential) equation describing populations of coupled oscillators and excitable elements~\cite{Kawamura,Kawamura2}, and also for fluid systems exhibiting stable oscillatory convection~\cite{Kawamura3,Kawamura4,Taira,Iima}. As we have formulated for the RD systems, we can derive the amplitude equation also for such systems, which can be used for analyzing their transient relaxation properties. 

In this chapter, we have presented the theory only for a single oscillatory system in the simplest lowest-order approximation with a single phase and a single amplitude.
Generalization to more complex cases is also possible along the lines presented in this chapter.
In general, even a single system may possess two or more phase variables when it has additional continuous translational symmetries e.g. in spatial directions~\cite{Kawamura4,Biktasheva} and possesses a torus solution rather than a limit-cycle solution. We may also need to consider two or more amplitude variables and consider the case of complex Floquet eigenvalues. 
Moreover, although the phase is decoupled from the amplitude at the lowest-order phase-amplitude description considered in this chapter, this is not the case if we proceed to the next order;
nonlinear phase-amplitude interactions can arise within a single system and, if coupled systems are considered, phase and amplitude interactions of three- or more systems can generally arise.
Such higher-order interactions can be a source of intriguing complex dynamics in nonlinear oscillatory systems; see, e.g., Refs.~\cite{Kuramoto2,Ermentrout2,Ashwin,Wilson2,Leon,Rosenblum,Wilson3,Kotani2}
for various types of higher-order descriptions and their consequences.

Macroscopic oscillatory systems in the real world are often made up of spatially distributed populations of interacting microsystems and modeled by PDEs including the RD equations.
Development of the phase-amplitude framework for such systems,
together with the recent advance in the Koopman operator approach to nonlinear dynamical systems~\cite{Mauroy,Mauroy2,KoopmanBook,Nakao2,Nakao3},
will provide us with a unified viewpoint and practical methods for the analysis, control, and design of such complex oscillatory systems.

\acknowledgements{
The author is grateful to A. Stefanovska and P. V. E. McClintock for invitation to write this chapter and for their kind advice. This work is financially supported by JSPS KAKENHI Grants JP17H03279, 18K03471, JP18H03287, and JST CREST JPMJCR1913.
}

\end{document}